\documentclass[reprint,aip]{revtex4-1}

\usepackage{physics}
\usepackage{amsmath}

\def\dbar{{\mathchar'26\mkern-12mu d}}
\begin{document}

\title{\centering{Entropy as an adiabatic invariant}}

\begin{abstract}
Note: This short article was submitted to Nature Physics as a
Correspondence.  The intention was to provide a brief albeit
significant criticism of the work of J. Dunkel and S. Hilbert,
\textit{Consistent Thermostatistics Forbids Negative Absolute
Temperatures}, Nature Physics \textbf{10}, (2014). The respected
editor decided not to publish the Correspondence.  We have
therefore decided to submit the paper to arXiv.
Comments/criticisms are welcomed, particularly from the authors of
the mentioned paper.
\end{abstract}

\maketitle

\textbf{To the editor ---} Recently Dunkel and Hilbert
\cite{dunkel2014} have argued that the consistency relation
$T(\pdv*{S}{A_\mu})_E=-(\pdv*{E}{A_\mu})_S=-\expval{\pdv*{H}{A_\mu}}$
favors the Gibbs entropy ($S_G$) over the Boltzmann entropy
($S_B$) since this relation is satisfied if and only if $S$ is an
adiabatic invariant. The first equality follows simply from
invertibility of entropy $S(E,A_\mu)\rightarrow E(S,A_\mu)$;
however, the second equality is rooted in assuming that a
mechanical adiabatic process is equivalent to a slow thermodynamic
adiabatic process. For a mechanical adiabatic process one can
write $\expval{\dd H/\dd t}=\dd E/\dd t=\sum_\mu
\expval{\pdv*{H}{A_\mu}}\dd A_\mu/\dd t$. On the other hand, for a
thermodynamic adiabatic process ($\dd S/\dd t=0$), one can write
$\dd E/\dd t=\sum_\mu (\pdv*{E}{A_\mu})_S  \dd A_\mu/\dd t$. The
equivalence of these two leads to the second relation in the
``consistency relation",
i.e.~$(\pdv*{E}{A_\mu})_S=\expval{\pdv*{H}{A_\mu}}$. As is
well-known $N$ is an important thermodynamic variable, which
contributes to the total energy through chemical work, $\; \dbar
W=\mu \dd N$, through the intensive parameter of chemical
potential, $\mu \equiv (\pdv*{E}{N})_S$. In fact in the
Supplementary Information, Dunkel and Hilbert explicitly mention
that $A_\mu$ can be an external parameter, \textit{``such as
volume, particle number, magnetic field strength, etc"}
\cite{dunkel2014}. Here we note that such a relation fails to hold
for the important parameter $A_\mu=N$ even when $S$ is chosen to
be the Gibbs entropy. In other words we show that thermodynamic
chemical potential $\mu_{th}=(\pdv*{E}{N})_S=-T(\pdv*{S}{N})_E$ is
not the same as ``mechanical" chemical potential
$\mu_m=\expval{\pdv*{H}{N}}$. This lack of consistency remains
true regardless of choice of entropy $S_G$ or $S_B$.  We therefore
argue that the consistency relation is invalid and should not be
used to argue in favor of one entropy ($S_G$) over the other
($S_B$).

To see that the consistency relation is invalid for $S=S_G=\ln
\lbrace \Tr[\Theta(E-H)]\rbrace$, for $ A_\mu=N$, one must simply
realize that the trace operation (or integration in phase space
for classical systems) is not interchangeable with differentiation
with respect to $N$, as the degrees of freedom of a quantum or
classical system (which determines the dimension of phase space in
classical system), crucially depends on $N$. This
interchangeability is assumed to always hold true (for any allowed
$A_\mu$) by Dunkel and Hilbert in order to prove the consistency
relation for $S_G$\cite{dunkel2014}. To see the above argument
more explicitly note that in the microcanonical ensemble,
$\expval{\pdv*{H}{N}}\equiv \Tr [\pdv*{H}{N}\delta
(E-H)]/\omega=-\Tr[\pdv*{\Theta(E-H)}{N}]/\omega$, where
$\omega=\Tr[\delta(E-H)]$. On the other hand,
$T_G(\pdv*{S_G}{N})_E=(\pdv*{S_G}{N})_E/(\pdv*{S_G}{E})_N=\pderivative*{N}
(\Tr [\Theta (E-H)])/\omega$. Therefore, one can see that the
consistency relation holds only if one can move in the
differentiation into the integration on state space, i.~e.~trace
operation. This is clearly not allowed and has in fact been proved
to be false in specific and explicit cases\cite{tavassoli2015}. If
one replaced $\Tr$ with $\int ...\int \dd ^{3N}p  \dd
^{3N}q/h^{3N}$ for a classical system, the lack of
interchangeability will become more evident.

As an example of why $\mu_{th} \neq \mu_m$, consider a simple
classical ideal gas. It is well-known that the chemical potential
of such a system is a large negative number
$\mu_{th}=(\pdv*{E}{N})_S<<0$ \cite{swendsen2012}. On the other
hand, the ``mechanical" chemical potential
$\mu_m=\expval{\pdv*{H}{N}}$ is clearly a nonnegative number as
one can imagine by measuring the change of total energy by adding
(a zero energy) particle.

We note that lack of validity of consistency relation is rooted in
the assumption that a slow thermodynamical adiabatic process is
equivalent to a mechanically adiabatic process, which is in turn
rooted in a mechanical interpretation of thermodynamics. A
mechanical adiabatic process is simply a slow process where $ \dd
A_\mu / \dd t$ is assumed to be very small. A thermodynamics
adiabatic (and quasi-static) process is one which entropy remains
constant. The main idea of Dunkel and Hilbert is that if entropy
is a mechanical adiabatic invariant then the two processes are
equivalent, i.e. $S=$const.  However, we have argued that Gibbs
entropy cannot be an adiabatic invariant with respect to $N$.
Therefore, since $N$ is a fundamental thermodynamic parameter,
arguments based on mechanical properties of Gibbs entropy such as
consistency and/or adiabatic invariance should be viewed with
extra care. It is important to note that since consistency
relation is in effect an equivalence relation between various
energy terms, such as mechanical work or chemical work, it is
important for all such terms to obey this equivalence. Therefore,
if such equivalence breaks down for a given term (i.e. with
respect to $N$), then the fundamental thermodynamic relation $\dd
E = T\dd S -p\dd V + \mu\dd N + ...$ cannot have a mechanical
analog even if $S$ is considered to be an adiabatic invariant with
respect to all but one energy term.  In other word, since all
systems of interest possess chemical potential, the lack of
equivalence in this quantity casts doubt on the reliability of
such consistency relation as a criterion to judge the validity of
various entropy definitions.

Next, we address some potential objections. The entire formulation
of the consistency relation is based on the microcanonical
(entropy) representation where one assumes that $S=S(E,V,N, ...)$.
Although the system is isolated or closed (i.e. fixed
$E,V,N,...$), one can clearly differentiate the entropy function
with respect to its independent variables,
i.e.~$(\pdv*{S}{N})_{E,V} = f(E,V,N)$. Otherwise, intensive
parameters such as temperature $T(E,V,N)=
\lbrace(\pdv*{S}{E})_{V,N}\rbrace^{-1}$ cannot be defined.
Furthermore, differentiation with respect to discrete variables is
a standard practice in thermodynamics \cite{swendsen2012}. Quantum
mechanics teaches us that energy is \emph{not} a continuous
variable. However, for large macroscopic systems (and even
otherwise), we perform partial differentiation with respect to
energy since $(\Delta E/E) \rightarrow 0$.  In much the same way,
particle number is considered a continuous variable since $(\Delta
N / N=1/N) \rightarrow 0$. Therefore, one cannot exempt $N$ from
the requirement of consistency relation, $A_\mu = N$.

We end by making a general comment about the ``mechanical
foundation of thermodynamics". Thermodynamics is a formal theory
based on lack of mechanical information about constituents of
macroscopic systems, which naturally leads to the existence (and
definition) of entropy.  Mechanics, on the other hand, assumes
availability of all relevant mechanical variables.  Clearly, these
two important theories are based on opposing set of assumptions.
The insistence to base thermodynamics on a mechanical foundation
has historically led to controversial issues, the most famous of
which are thermodynamic irreversibility vs.~mechanical
reversibility as well as the problem of Maxwell's
demon\cite{swendsen2012}. The work of Dunkel and Hilbert yet
provides another example of how such an insistence leads to
controversial and inconsistent results. We finally emphasize that
our results should not be interpreted as favoring one entropy
definition over another, but a critique of the so called
``consistency" relation which equates mechanical and
thermodynamical adiabatic processes. \setlength{\parindent}{0pt}

\rule{\linewidth}{1pt}
\textbf{Afshin Montakhab* and Arash Tavassoli}
\setlength{\parskip}{0cm plus0mm minus0mm}

Department of Physics, College of Sciences, Shiraz University, Shiraz 71946-84795, Iran

*email: montakhab@shirazu.ac.ir

\end{document}